\documentclass[12pt]{iopart}

\usepackage{graphicx}
\usepackage{iopams}
\usepackage{float}

\usepackage{hyperref}
\usepackage{caption}

\begin{document}
\title[Arm locking performance with the new LISA design  ]{Arm locking performance with the new LISA design}
\author{Sourath Ghosh$^1$, Josep Sanjuan $^2$, Guido Mueller$^3$}

\address{$^{1,2,3}$Department of Physics, University Of Florida, USA}

\ead{s.ghosh@ufl.edu}
\vspace{10pt}
\begin{indented}
\item[]November 2021
\end{indented}

\begin{abstract}
The Laser Interferometer Space Antenna (LISA) is a future space-based gravitational wave (GW) detector designed to be sensitive to sources radiating in the low frequency regime (0.1\,mHz to 1\,Hz). LISA's interferometer signals will be dominated by laser frequency noise which has to be suppressed by about 7 orders of magnitude using an algorithm called Time-Delay Interferometry (TDI). Arm locking has been proposed to reduce the laser frequency noise by a few orders of magnitude to reduce the potential risks associated with TDI. In this paper, we present an updated performance model for arm locking for the new LISA mission using $2.5~{\rm Gm}$ arm lengths, the currently assumed clock noise, spacecraft motion based on LISA Pathfinder data and shot noise. We also update the Doppler frequency pulling estimates during lock acquisition.  
\end{abstract}

\section{Introduction}
\label{sec:Introduction}
The Laser Interferometer Space Antenna (LISA)~\cite{L3proposal} will be the
first dedicated space-based gravitational wave observatory. It is
scheduled to be launched in the 2030s and will target the $0.1\,{\rm mHz}$
to $1\:{\rm Hz}$ frequency range, opening a window that is believed
to be unobservable with current and future ground-based observatories
such as LIGO~{\cite{LIGOScientific:2014pky,LIGOScientific:2007fwp}}, VIRGO~{\cite{VIRGO:2012dcp,VIRGO:2014yos}}, KAGRA~{\cite{PhysRevD.88.043007,akutsu2020overview}}, GEO600~{\cite{Grote:2010zz,Dooley_2016}}, ET~{\cite{Kroker:2015pmg}} or the Cosmic
Explorer~{\cite{reitze2019cosmic}}.
LISA consists of three spacecraft oriented in an approximate equilateral triangle with $2.5~{\rm Gm}$ arm length. Each spacecraft hosts a pair of free-falling test masses and a pair of lasers inside a movable optical sub-assembly (MOSA). Hence, there are two laser links between each pair of spacecraft (six inter-spacecraft links in total). In order to detect GW with characteristic strain of $\mathcal{O}(10^{-21})$ , all inter-test mass optical path lengths must ultimately be measured to picometer accuracy.

LISA will not only be the largest laser interferometer ever built
but also the one with the largest arm length difference of up to $35\,000\,{\rm km}$.
It is also one of the most dynamic interferometer with relative test
mass velocities of around $\pm10\,{\rm m/s}$ resulting in Doppler shifts
of up to $\pm10\,{\rm MHz}$ for the $1.064\,\mu{\rm m}$ laser
wavelength~\cite{L3proposal}. As with all interferometers, the arm length difference
increases the interferometer's susceptibility to laser frequency noise,
which LISA overcomes by employing Time-delay Interferometry (TDI)
{\cite{10.1007/s41114-020-00029-6}}. TDI uses linear combinations of time
shifted interferometer signals to synthesize an artificial (near-) equal
arm interferometer in post-processing. 

The first generation TDI (TDI 1.0)  assumed a static arm length
mismatch \cite{PhysRevD.65.082003}. This led to simple requirements on the allowed residual
frequency noise, $\delta\nu$, for a given uncertainty, $\Delta L$, of the arm length mismatch:
\begin{equation}
\delta\nu=\nu\frac{\delta l}{\Delta L}\label{eq:dl_dnu}
\end{equation}
where $\delta l$ is the interferometer sensitivity. In TDI, $\Delta L$
translates into an uncertainty of how much the interferometer signals
have to be time shifted to cancel laser frequency noise. Soon after
it was realized that changes in $\Delta L$  during the $8.3\,{\rm seconds }$
light travel time in each arm add to the uncertainty in $\Delta L$ \cite{10.1007/s41114-020-00029-6}
and tighten the resulting laser frequency noise requirement in terms of amplitude spectral density (ASD):
\begin{equation}
\widetilde{\delta\nu}_{\rm TDI\,1.0}(f)<1.7\times\sqrt{1+\left(\frac{2.8\,{\rm mHz}}{f}\right)^{4}}\,\frac{{\rm Hz}}{\sqrt{{\rm Hz}}},
\end{equation}
which is challenging to guarantee using a local frequency reference. LISA is expected to use the second generation TDI (TDI 2.0) \cite{10.1007/s41114-020-00029-6}, which
uses essentially two time-shifted TDI 1.0 combinations to compensate
for the arm length changes during the light travel time. The resulting
requirement on laser frequency noise for TDI 2.0 is:
\begin{equation}
\widetilde{\delta\nu}_{\rm TDI\,2.0}(f)<282\times \sqrt{1+\left(\frac{2.8\,{\rm mHz}}{f}\right)^{4}}\,\frac{{\rm Hz}}{\sqrt{{\rm Hz}}}
\end{equation}
assuming a ranging uncertainty of $\Delta L<1\:{\rm m}$. 

\begin{figure}[H]
 \centering
 \includegraphics[scale=0.4]{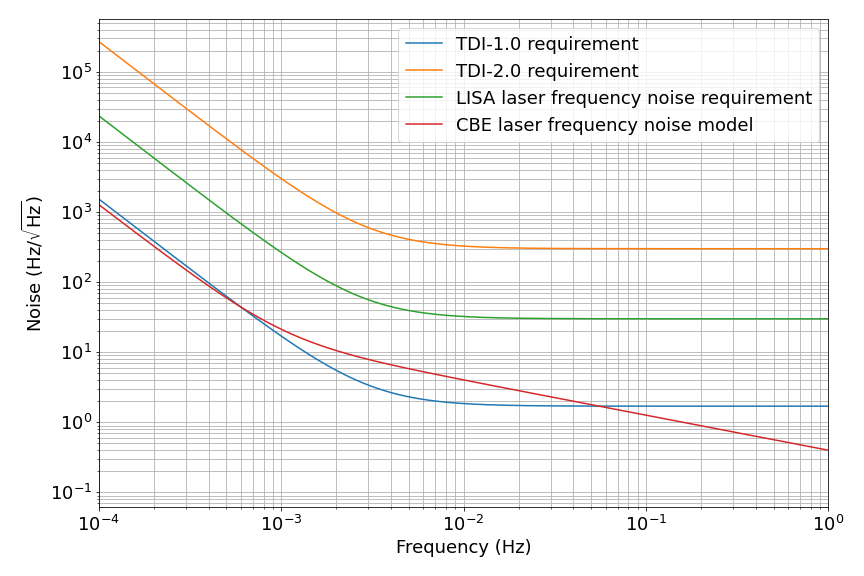}
\caption{TDI, LISA laser frequency noise requirements and CBE laser noise model}
\label{fig:TDIvsinputlasernoise}
\end{figure}

The current laser frequency noise requirement has been set to \cite{freqplanning}:
\begin{equation}
\widetilde{\delta\nu}_{\rm Req.}(f)<30\times\sqrt{1+\left(\frac{2.8\,{\rm mHz}}{f}\right)^{4}}\,\frac{{\rm Hz}}{\sqrt{{\rm Hz}}}    
\label{eqn:Lisareq}
\end{equation}
to keep some margin while the current best estimate (CBE)~\cite{2017}
for the laser frequency stabilization system
is:

\begin{equation}
\widetilde{\delta\nu}_{\rm CBE}(f)\approx0.4\left(\frac{f}{1\,{\rm Hz}}\right)\left(1+\left(\frac{1\,{\rm mHz}}{f}\right)^2\right)^{3/4}\frac{{\rm Hz}}{\sqrt{{\rm Hz}}}\,.
\end{equation}
The requirements for TDI 1.0, TDI 2.0, the LISA requirement and
the CBE are plotted in Fig~\ref{fig:TDIvsinputlasernoise}. While the CBE nearly
meets the TDI 1.0 requirements, the project requirement $\widetilde{\delta\nu}_{\rm Req.}$
requires to use TDI 2.0. In both cases, LISA relies on seven to eight
orders of common mode rejection of laser frequency noise to reach
design sensitivity.

Arm locking \cite{Sheard:2003fz} has been proposed to further reduce the
laser frequency noise by using the LISA arms as frequency references.
Within the LISA measurement band, the LISA arms are, by virtue of their
free falling end points the best frequency references available. However,
the long light travel times and the relative motion of the end points
require tailored sensor and controller designs which have been developed
during the first decade of this century for the original $5\,{\rm Gm}$
LISA mission. This has been summarized very well in {\cite{Kirk,Yinan}}. In this paper, we apply the lessons learned to the new LISA
design with its shorter arms, different orbits, improved clock noise,
and LISA Pathfinder (LPF) based model for the residual spacecraft motion to supply a baseline for follow up studies. \\

The paper is organized as follows: Section \ref{sec:Arm locking overview and challenges} briefly reviews arm locking, the challenges involved in the arm locking sensor and controller design and how clock noise, shot noise and residual spacecraft motion is included in the model (see \cite{Kirk,Yinan} for a detailed study). Section \ref{sec:LISA mission parameters} discusses the new LISA mission parameters and noise ASDs  that are relevant to the arm locking performance. In addition, changes in the controller design and the averaging time for the initial Doppler estimates are presented considering the new LISA parameters. Section \ref{sec:Modified dual arm locking noise performance for the new LISA mission}  contains the calculations for the new LISA arm locking noise performance and the Doppler frequency pulling transient. Section \ref{sec:conclusion} discusses the implications and future prospects.

\section{Arm locking review}
\label{sec:Arm locking overview and challenges}
\begin{figure}[H]
     \centering
     \begin{minipage}{0.58\textwidth}
     \includegraphics[width=\textwidth]{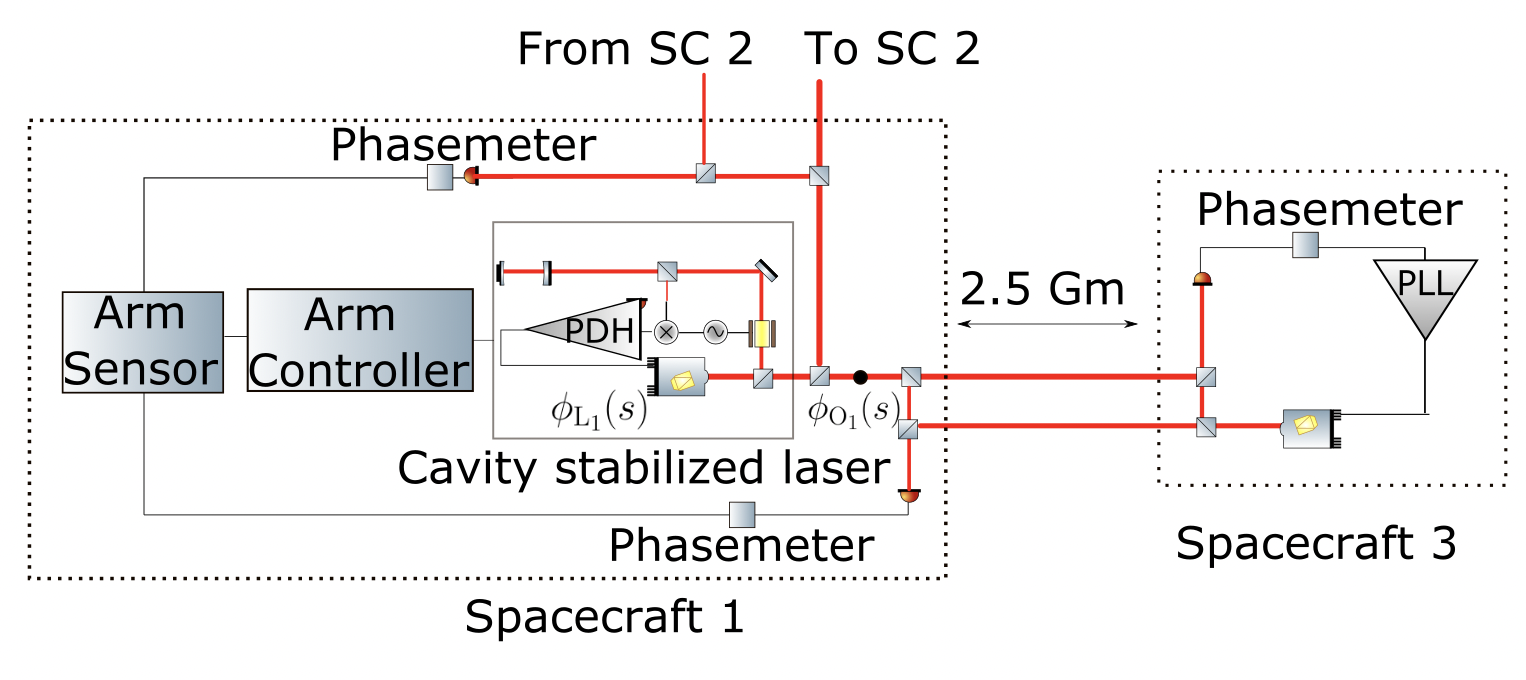}
    \end{minipage}
    \hfill
    \begin{minipage}{0.4\textwidth}
    \includegraphics[width=\textwidth]{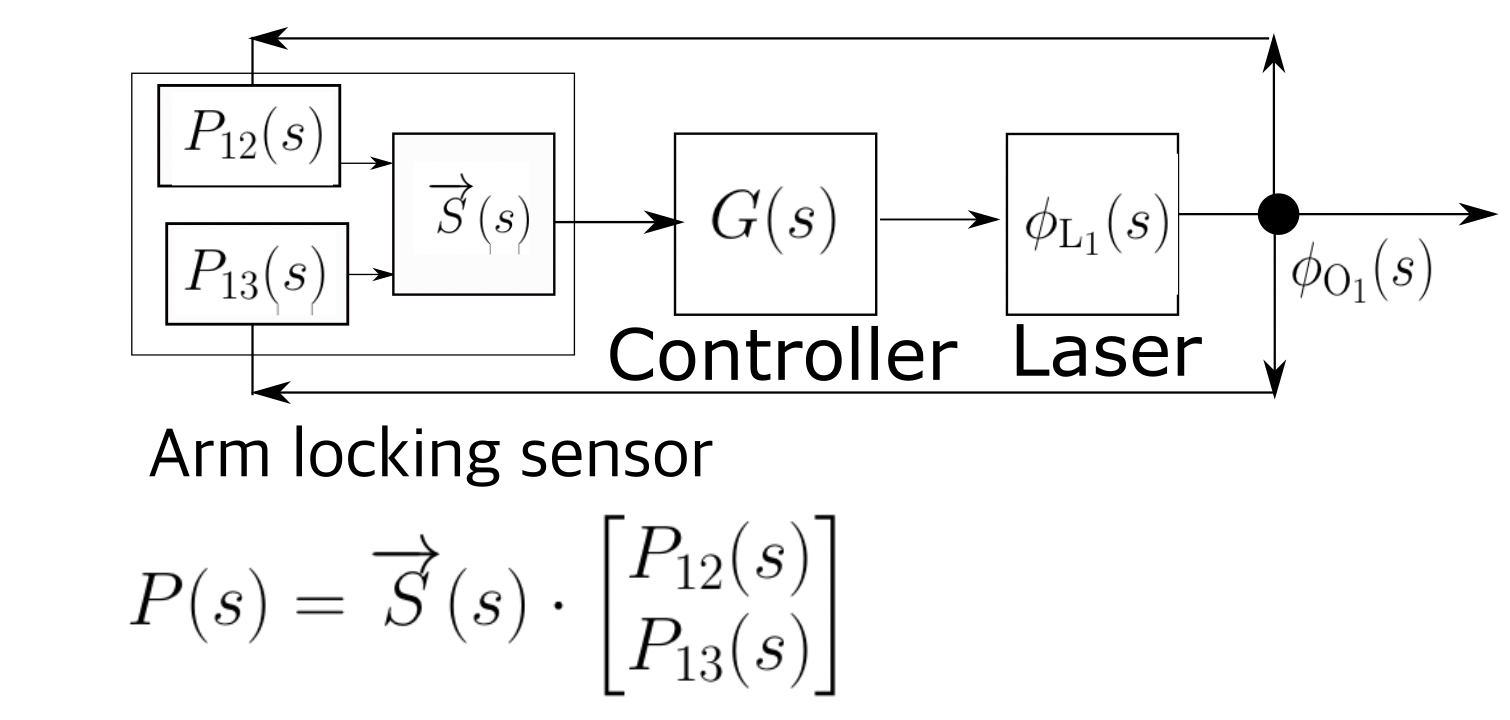} 
    \end{minipage}
    \caption{Arm locking scheme (left) and equivalent control loop diagram (right)}
    \label{fig:Arm locking }
\end{figure}

\begin{figure}[H]
\centering
\includegraphics[scale=0.4]{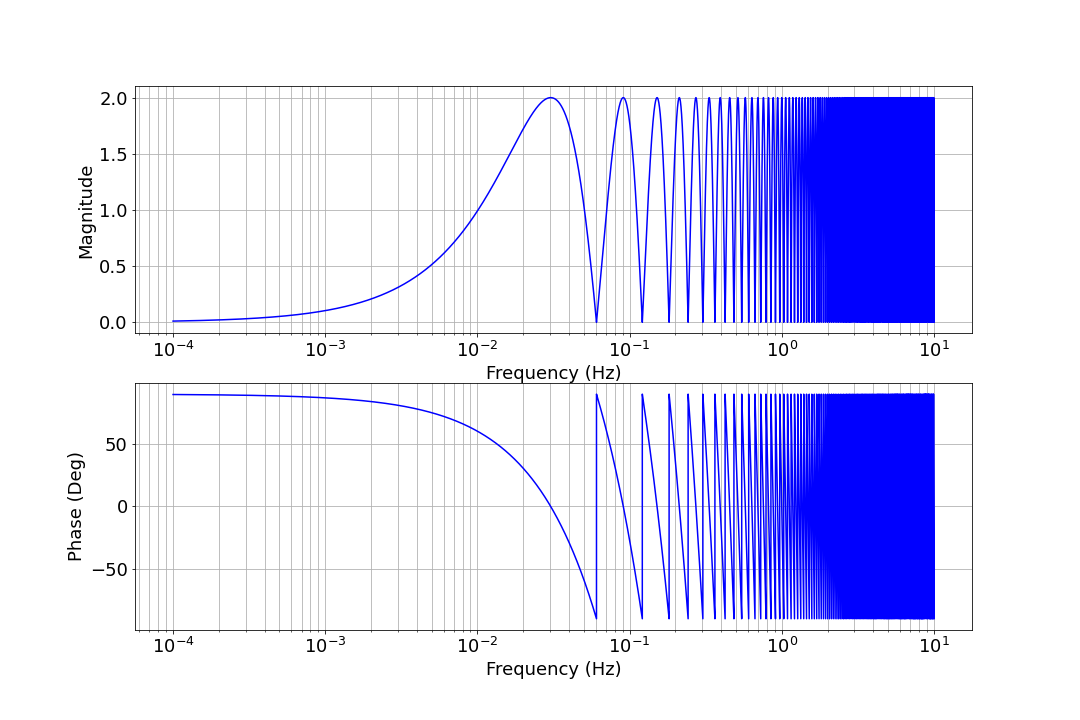}
\label{fig:singlearmsensorphase}
\caption{Bode plot of the single arm sensor.}
 \label{fig:Singlearmbodeplot}
\end{figure}

Arm locking was originally proposed by  Sheard et al. in 2003 \cite{Sheard:2003fz}). The idea was to take advantage of the phase lock loop (PLL) between two spacecraft and use the inter-spacecraft phasemeter measurement as feedback to correct the laser frequency. This is sketched in FIG [\ref{fig:Arm locking }] where the frequency of the laser on spacecraft\,1 is locked to the distance between spacecraft\,1 and spacecraft\,3. This frequency stabilization scheme, now known as single arm locking, is essentially a Mach-Zehnder frequency stabilization scheme with one sub-meter scale arm on the optical bench and one very long arm ($\tau_{13}\approx 16.6~{\rm seconds}$, round trip) which also changes its length at a rate of a few meter per second but is extremely stable in the LISA band. \\\\
The noise suppression achieved  is given by the magnitude of the closed loop transfer function in the Laplace domain $T_{{\rm CL}}(s)$ (where $s=i\omega=2\pi i f$): 

\begin{equation}
    \vert T_{\rm CL}(s) \vert=\left|\frac{\phi_{{\rm O}1}(s)}{\phi_{{\rm L}1}(s)}\right|=\left| \frac{1}{1+G(s)P_{13}(s)}\right|
\label{eqn:Single arm noise supression}
\end{equation}
where $P_{13}(s)=1-\exp{(-s \tau_{13})}$ is the arm transfer function and $G(s)$ is the arm locking controller.\\\\
From eqn \ref{eqn:Single arm noise supression}, it is evident high controller gain is desired throughout the LISA science band for achieving high noise rejection. From the Bode plot (Fig \ref{fig:Singlearmbodeplot}), we see that the single arm transfer function has zero response at DC and multiples of the free spectral range $c/2L$. Additionally, these null frequencies have phase discontinuities where the phase changes from $\pi/2$ to $-\pi/2 $ rad. Together with any suitable controller, the open loop transfer function $G(s)P_{13}(s)$ will cross unity gain just before and after each null within the nominal controller bandwidth (e.g. frequencies where $G(s)>1$). The additional frequency dependency of the controller adds an additional phase loss which can lead to unity gain oscillations if the phase loss reaches $180$ degrees. The solution first proposed in \cite{Sheard:2003fz} is to use a controller with a slope of $G(s)=\frac{1}{\sqrt{s}}$ in order to maintain a phase margin from $180$ degrees at these unity gain frequencies. 


In addition to the constraints on the controller slope, the controller gain is also restricted by the Doppler frequency pulling. Relative motion between spacecraft imparts a Doppler frequency shift on the transponded beam.  This Doppler frequency shift cannot be estimated and subtracted accurately in real time. In order to maintain the desired heterodyne beatnote frequency, the Doppler estimation error is compensated by changing the local laser frequency. Hence, the Doppler error adds to the laser frequency after every round trip  light travel time. In other words, for single arm locking the drift rate $\left(\frac{d \nu_{{\rm CL}}}{dt}\right)$ of the laser frequency scales by the Doppler error $\Delta \nu_{\rm DE}$ and the inverse of the round trip light travel time:

\begin{equation}
    \frac{d\nu_{{\rm CL}}(t)}{dt}= \frac{\Delta \nu_{\rm DE}}{\tau_{13}}Y_{G_{{\rm single}}}(t)
\label{eqn:Dopplersinglearmsimple}
\end{equation}
 where $Y_{G_{\rm single}}(t)$ is a controller dependent function.
Large drifts in this local laser frequency (and consequently the transponder laser frequencies) will lead to  mode-hops and failures of the frequency control system. \\\\


\subsection{Arm locking sensor design}
\begin{figure}[H]
     \centering
\includegraphics[scale=0.4]{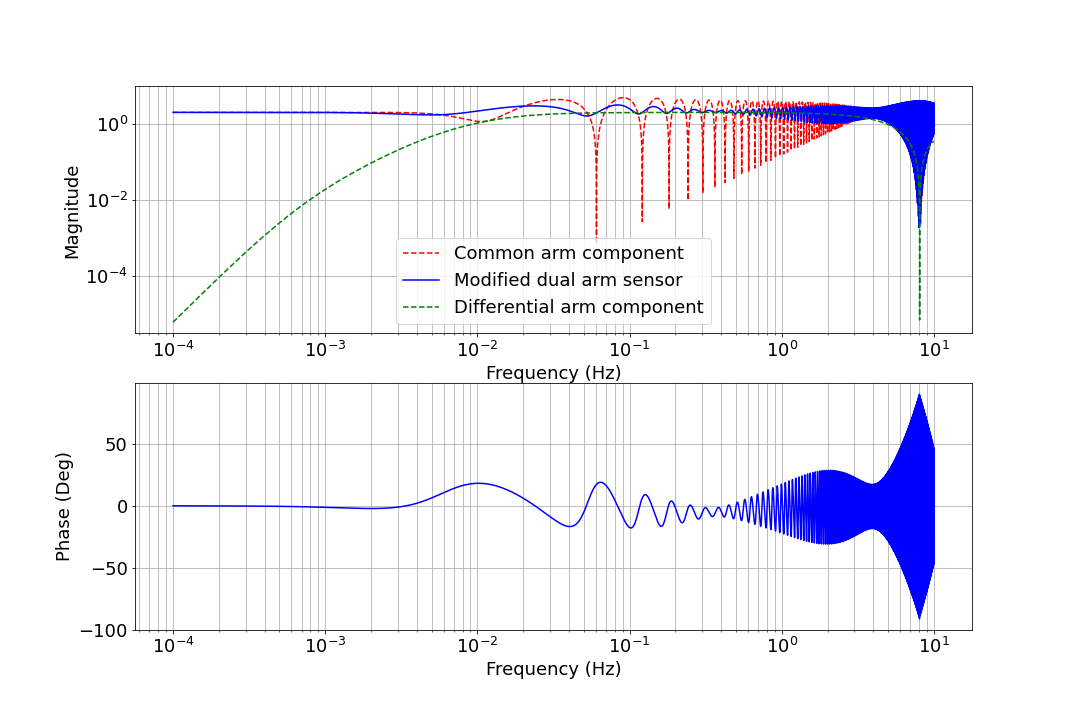}
    \caption{Bode plots for the modified dual arm locking sensor with the arm length mismatch of  $\Delta \tau= 62~{\rm  ms}$. The magnitude plot also includes contributions of the common and differential arm components.}
    \label{fig:MDAL_sensor}
\end{figure}

Following the work by Sheard et al. \cite{Sheard:2003fz}, there have been several works (\cite{Sutton} \cite{Yinan},\cite{Kirk}) on optimizing arm locking by constructing a sensor signal using frequency dependent linear combinations of signals from both arms. This resulted in sensor signals that allow for more aggressive controllers. We continue on the work done by McKenzie et al. \cite{Kirk} and use the modified dual arm locking (MDAL) sensor, which can be expressed as a combination of the common and differential arm transfer function:

\begin{equation}
    P_{\rm M}(s)=P_{+}(s)H_{+}(s)+P_{-}(s)H_{-}(s)
\label{eqn:MDAL_sensor}    
\end{equation}
where $P_{+}(s)=P_{12}(s)+P_{13}(s)$ and $P_{-}(s)=P_{12}(s)-P_{13}(s)$ are the common and differential arm transfer functions with $H_{+}(s)$ and $H_{-}(s)$ being their respective frequency dependent weight factors (see section \ref{subsec:Modified dual arm  sensor parameters}).
Figure \ref{fig:MDAL_sensor} contains the Bode plot of this sensor along with the common and differential arm components. We see that  the sensor nulls are pushed to higher frequencies which guarantees no unity gain frequencies in the LISA science band and allows the use of  a far more aggressive controller (compared to single arm locking) at low frequencies to ramp up the low frequency gain.\\\\
Additionally, we note that the sensor magnitude is dominated by the common arm at low frequencies and the differential arm has zero response at DC. This ensures that the Doppler frequency pulling scales  as the inverse of the average round trip light travel time $1/\bar{\tau}$ as opposed to the inverse of the differential time delay $1/\Delta \tau$ for the differential arm sensor. Similar to Eq. \ref{eqn:Dopplersinglearmsimple} the Doppler frequency pulling rate for modified dual arm locking is :
\begin{equation}
    \frac{d\nu_{{\rm CL}}(t)}{dt}\approx \frac{\Delta \nu_{\rm DE}}{\bar{\tau}}Y_{{\rm G}_{\rm MDAL}}(t)
\label{eqn:MDALsinglearmsimple}
\end{equation}
$Y_{{\rm G}_{\rm MDAL}}(t)$ is a controller dependent function. As done in \cite{Kirk} we use an AC coupled controller which introduces damping in $Y_{{\rm G}_{\rm MDAL}}(t)$ 
    


\subsection{Doppler frequency pulling at lock acquisition and arm locking controller design}

\begin{figure}[H]
     \centering
     \includegraphics[scale=0.6]{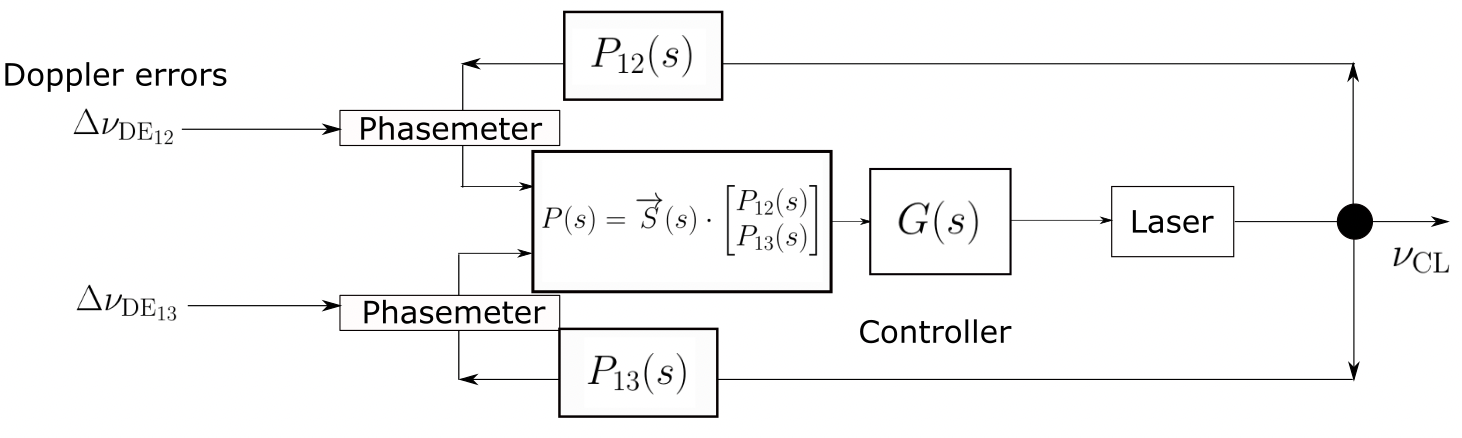}
         \caption{Schematic of the Doppler frequency error in the modified dual arm locking control loop.}
         \label{fig:MDAL_dopppler_control_loop}
\end{figure}

In this section, we study the Doppler frequency pulling response at lock acquisition since it further restricts the controller design. Figure \ref{fig:MDAL_dopppler_control_loop} shows the block diagram of the system. The Doppler errors for both arms, $\nu_{{\rm DE}12}$ and $\nu_{{\rm DE}13}$, enter at both phasemeter channels and lead to the laser frequency pulling response within the modified dual arm sensor.



Before arm locking is engaged, a measurement of the Doppler shift and its derivatives will be made from the  averaged phasemeter readout. At lock acquisition, these estimates will be subtracted out from the phasemeter measurement.  Hence, the measurement errors in the initial zeroth, first, and second Doppler derivatives (henceforth denoted by $\nu_{0},\gamma_{0} \,{\rm  and }\,\alpha_{0}$, respectively) lead to a time varying  Doppler error 
\begin{equation}
   \Delta \nu_{{\rm DE}ij}(t)\approx\nu_{0ij}+\gamma_{0ij}t+\frac{\alpha_{0ij}}{2}t^2 .
\end{equation}
The Doppler pulling of the laser frequency  at lock acquisition is given by the step response of the arm locking control loop. This response is split as a sum of the three Doppler derivative contributions and is given by:

\begin{eqnarray}
\nu_{\rm CL}(t)=&\mathcal{L}^{-1}\left(\frac{G(s)}{(1+G(s)P_{\rm M}(s))s}\left(H_{+}(s)\nu_{0+}+H_{-}(s)\nu_{0-}\right)\right)\label{eqn:Dopplerpul3derivatives}\\\nonumber
&+\mathcal{L}^{-1}\left(\frac{G(s)}{(1+G(s)P_{\rm M}(s))s^2}\left(H_{+}(s)\gamma_{0+}+H_{-}(s)\gamma_{0-}\right)\right)\\\nonumber
&+\mathcal{L}^{-1}\left(\frac{G(s)}{(1+G(s)P_{\rm M}(s))2s^3}\left(H_{+}(s)\alpha_{0+}+H_{-}(s)\alpha_{0-}\right)\right)
\end{eqnarray}\\
where $\mathcal{L}^{-1}$ denotes the inverse Laplace transform and the subscripts $'+'$ and $'-'$ represent the common and differential Doppler derivative errors:
\begin{eqnarray}
    \nu_{0\pm}=&\nu_{12}\pm \nu_{13}\\
    \gamma_{0\pm}=&\gamma_{12}\pm \gamma_{13}\\
    \alpha_{0\pm}=&\alpha_{12}\pm \alpha_{13}
\end{eqnarray}
Equation \ref{eqn:Dopplerpul3derivatives} shows that using an ideal constant infinite gain controller, would lead to a steep power-law ramp-up in the laser frequency $\mathcal{O}({\rm GHz})$ in the first $12$ days assuming Doppler estimates corresponding to $200$ seconds averaging window (table \ref{tab:Initial errors in Doppler estimates and its derivatives})). In order to dampen this frequency pulling the controller used in \cite{Kirk} is AC coupled with a series of high pass filters at frequencies below the LISA band. We use a controller with the same parametric form but tailor the poles and zeros to optimize the noise performance and the Doppler pulling for the new LISA mission (see section \ref{subsec:Controllerpolesandzeros})\\\\
In summary, the controller must be i) AC coupled in order to dampen the Doppler frequency pulling; ii) Have high gain in the LISA band to achieve maximum noise suppression; iii) Have a roll off at frequencies above the LISA band that ensures maximum bandwidth by avoiding unity gain frequencies when the  phase is close to $180 ^{\circ}$.\\

\subsection{Other noise sources}

\begin{figure}[H]
     \centering
     \includegraphics[width=\textwidth]{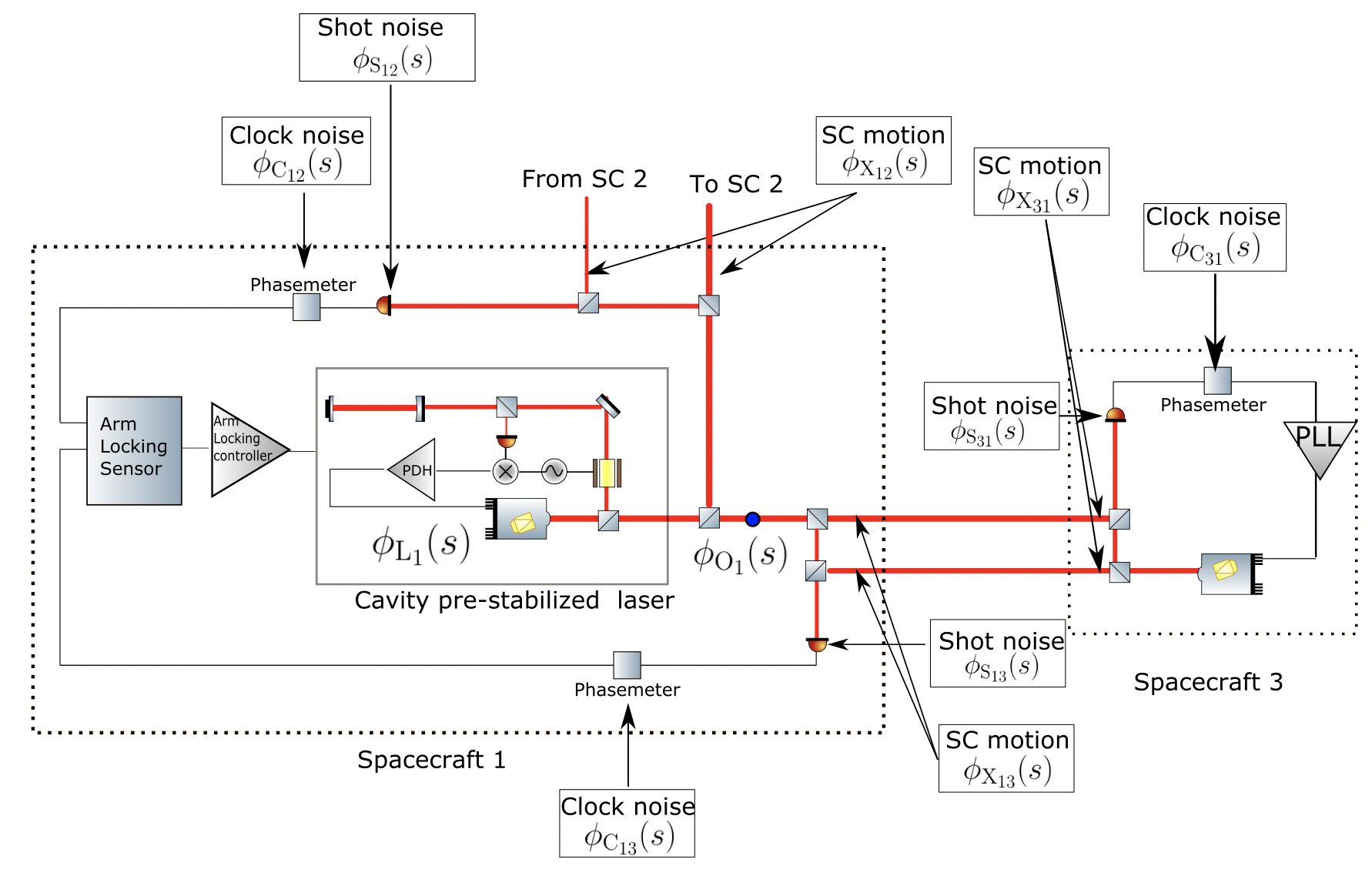}
         \caption{Arm locking control loop including clock noise, shot noise and spacecraft motion contributions}
         \label{fig:Arm locking loop with noise terms}
\end{figure}

In addition to the laser's intrinsic frequency noise we need to account for other noise sources that contribute to the sensor signals. As a result of not being suppressed in loop, these noise sources ultimately place a lower bound on the arm locking noise performance. The dominant  noise sources are clock noise, shot noise, and noise induced from spacecraft motion (see Fig~\ref{fig:Arm locking loop with noise terms} ).

Solving the arm locking loop dynamics \cite{Kirk}, we can calculate the stabilized output frequency noise in terms of the arm locking sensor, arm locking controller, the input laser frequency noise ($\phi_{{\rm L}1}(s)$) and the external noise contributions:

\begin{eqnarray}
    \phi_{{\rm O}_1}(s)=&\frac{\phi_{{\rm L}_1}(s)}{1+G(s)P_{\rm M}(s)}-\frac{G(s)}{1+G(s)P_{\rm M}(s)}\overrightarrow{S}\cdot \left(\overrightarrow{N_{\rm C}}+\overrightarrow{N_X}+\overrightarrow{N_S}\right)
\label{eqn:armlockloopoutallnoises}
\end{eqnarray}\\
where 
\begin{eqnarray}
    \overrightarrow{N_S}&=&\left[\matrix{\phi_{{\rm S}_{12}}(s)+\phi_{{\rm S}_{21}}(s)e^{-s \tau_{21}}\cr \phi_{{\rm S}_{13}}(s)+\phi_{{\rm S}_{31}}(s)e^{-s \tau_{31}}\cr}\right]\\
    \overrightarrow{N_{\rm C}}&=&\left[\matrix{\phi_{{\rm C}_{12}}+\phi_{{\rm C}_{21}}(s)e^{-s \tau_{21}}\cr
    \phi_{{\rm C}_{13}}+\phi_{{\rm C}_{31}}(s)e^{-s \tau_{31}}}\right]\\
    \overrightarrow{N_X}&=&\left[\matrix{-\phi_{{\rm X}_{12}}(1+e^{-s \tau_2})-2\phi_{{\rm X}_{21}}(s)e^{-s\tau_{21}}\cr-\phi_{{\rm X}_{13}}(1+e^{-s \tau_3})-2\phi_{{\rm X}_{31}}(s)e^{-s\tau_{31}}}\right]\\
    \overrightarrow{S}&=&\left[\matrix{H_{+}(s)\cr H_{-}(s)}\right]
\end{eqnarray}
 $\phi_{{\rm S}_{ij}},\phi_{{\rm C}_{ij}}\,{\rm  and }\,\, \phi_{{\rm X}_{ij}}$ are the (mutually independent) shot noise, clock noise and spacecraft motion terms coupled to laser frequency of the beam travelling from ${\rm SC}_{i}$ to ${\rm SC}_{j}$.\\
 
\section{New LISA mission parameters and noise ASDs}\label{sec:LISA mission parameters}
In this section we specify new LISA mission parameters, noise ASDs, and initial Doppler estimation errors that we assume for computing the arm locking noise performance and the Doppler frequency pulling using Eqs. \ref{eqn:armlockloopoutallnoises} and \ref{eqn:Dopplerpul3derivatives} respectively.
\subsection{Modified dual arm  sensor parameters}
\label{subsec:Modified dual arm  sensor parameters}
The modified dual arm sensor is constructed as a combination of the common and differential arm sensor as shown in Eq.\ref{eqn:MDAL_sensor}. All four terms in this equation are characterized by
the average round trip light travel time ($\bar{\tau}$) and the differential light travel time ($\Delta \tau$). The frequency-dependent coefficients are constructed out of the  filters given in table \ref{tab:Filters that enter the modified dual arm locking sensor}\footnote{Note that these definitions  are equivalent to the ones used by \cite{Kirk}}:


\begin{eqnarray}
    \fl H_{+}(s)&=&H_{+{\rm LPF}}(s)+H_{+{\rm HPF}}(s)=\left(\frac{g_ag_b(s+z_b)}{s(s+p_b)}\right)+\left(\frac{s^4}{(s+p_c)(s+p_d)(s+p_e)^2}\right)\\
    \fl H_{-}(s)&=&\left(\frac{g_f g_g s^4(s+z_g)}{s(s+p_c)(s+p_d)(s+p_e)^2}\right)\label{eqn:diffintegrator}
\end{eqnarray}
\begin{table}[H]
\begin{center}
\scalebox{0.9}{
    \begin{tabular}{|c|c|c|c|}
    \hline
        Filter & Zeroes & Poles & Gain \\
        \hline

        \hline
        $H_{+LPF}(s)$& $z_b=2\pi \times 5 /(13 \bar{\tau})$& $
        p_{a}=0, \; 
        p_{b}=2\pi\times 5 /(2\bar{\tau})$& $g_a=\bar{\tau}^{-1},\: g_b=p_b/z_b$\\
        \hline
        $H_{+HPF}(s)$&$z_c=0,z_d=0,z_e=0$&$p_c=7/(5\bar{\tau}),p    _{d}=11/(20\bar{\tau}),p_{e}=2\pi/(90             \bar{\tau})$&$g_c=g_d=g_e=1$\\
        \hline
                $H_{-}(s)$ &$z_c,z_d,z_e,$$z_g= 2\pi \times 10/ \Delta \tau$ &$p_c,p_d,p_e,p_f=0,$ &$g_f=\Delta \tau^{-1}$\\
        
        &   & $p_g= 2\pi\times0.1394/\Delta \tau$,$p_h= 2\pi \times 5 /(2\Delta \tau)$&$g_g=p_gp_h/z_g$\\
        \hline
    \end{tabular}}
\caption{Filters that enter the modified dual arm locking sensor}
\label{tab:Filters that enter the modified dual arm locking sensor}
\end{center}
\end{table}

\begin{figure}[H]
\centering
\includegraphics[scale=0.4]{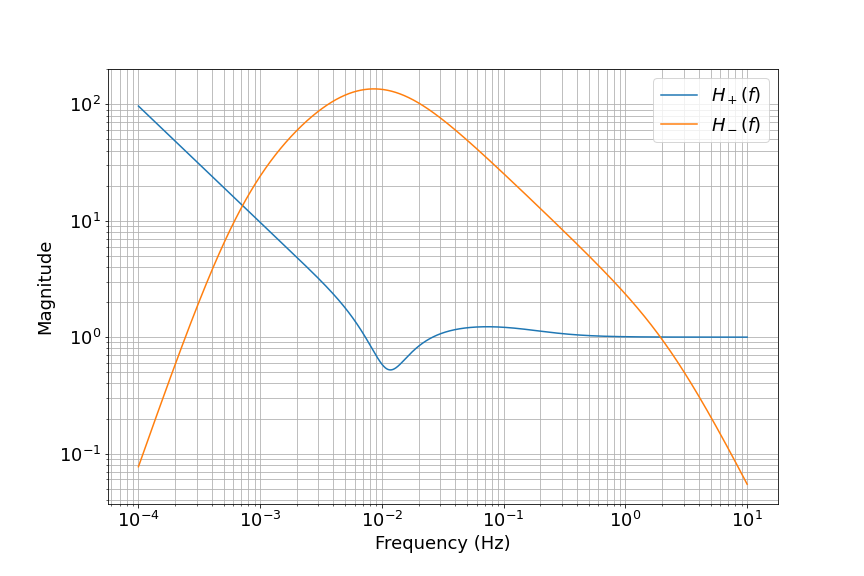}
\caption{Frequency dependent weighting factors of the common and differential arm sensor for modified dual arm locking.}
\label{fig:Hplusminus}
\end{figure}

$H_{+}$ is essentially constructed as a sum of a low pass $H_{+{\rm LPF}}$ and a unity gain high pass filter $H_{+{\rm HPF}}$ while $H_{-}$ is essentially a  band pass filter as shown in Fig. \ref{fig:Hplusminus}. \\\\

\begin{figure}[H]
        \centering
            \includegraphics[scale=0.4]{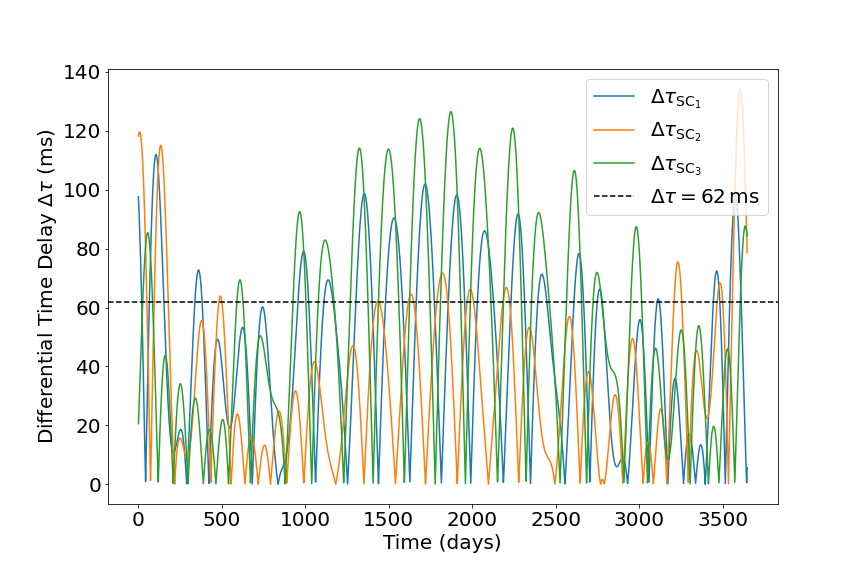}
            \caption{Variation in $\Delta \tau$ over mission timescale. The blue, red and green traces correspond to cases with $
            {\rm SC}_1$, ${\rm SC}_2$ and ${\rm SC}_3$ being primary spacecraft, respectively. The plots are derived from the orbit simulation \cite{Orbit}.}
            \label{fig:deltataumissiontimescale}
    \end{figure}

The quantities $\bar{\tau}$ and $\Delta \tau$ dynamically change over the mission duration Fig~\ref{fig:deltataumissiontimescale}). For our analysis we choose $\bar{\tau}=16.6 ~{\rm seconds}$ and $\Delta \tau =62~{\rm ms}$ which are nominal values over the mission duration.

\subsection{Noise ASDs}
In this sub-section, we give an overview and approximate ASDs of the significant noise sources that enter the arm locking loop. The input cavity stabilized laser frequency noise is assumed to be the  LISA requirement curve (i.e. $\phi_{\rm L1}(f)=\widetilde{\delta} \nu_{req}(f)$ in Eq.\ref{eqn:Lisareq}).\\

From \cite{Clocknoisereference}, the   current value of the fractional frequency fluctuations of the USO is given by 
\begin{equation}
    y(f)=8.2\times 10^{-14}\sqrt{\frac{1\,{\rm Hz}}{f}}\frac{1}{\sqrt{{\rm Hz}}}
\end{equation}
which is $1.5$ orders of magnitude better than the value assumed for the original LISA mission. The clock induced laser frequency noise scales with the beatnote frequency $(\Delta_{ij})$ and is given by:
\begin{equation}
    \phi_{{\rm C}ij}(f)=\Delta_{ij}y(f)
\end{equation} 
We assume $\Delta_{12}=14 ~ {\rm  MHz}$ and $\Delta_{13}=15~ {\rm  MHz}$ as the "worst case" scenario for computing the net clock noise vector $N_{c}$ in Eq.\ref{fig:Arm locking loop with noise terms}\\

The shot noise is given in terms of incident light power $P_{D}$ by:
\begin{equation}
    \phi_{{\rm S}_{ij}}(f)=\sqrt{\frac{(hc/\lambda)}{P_{d}}}\left(\frac{f}{1\,{\rm Hz}}\right)\frac{{\rm Hz}}{\sqrt{{\rm Hz}}}=10\left(\frac{f}{1\,{\rm Hz}}\right)\frac{\mu{\rm Hz}}{\sqrt{{\rm Hz}}}
\end{equation}

\begin{figure}[H]
        \centering
            \includegraphics[scale=0.4]{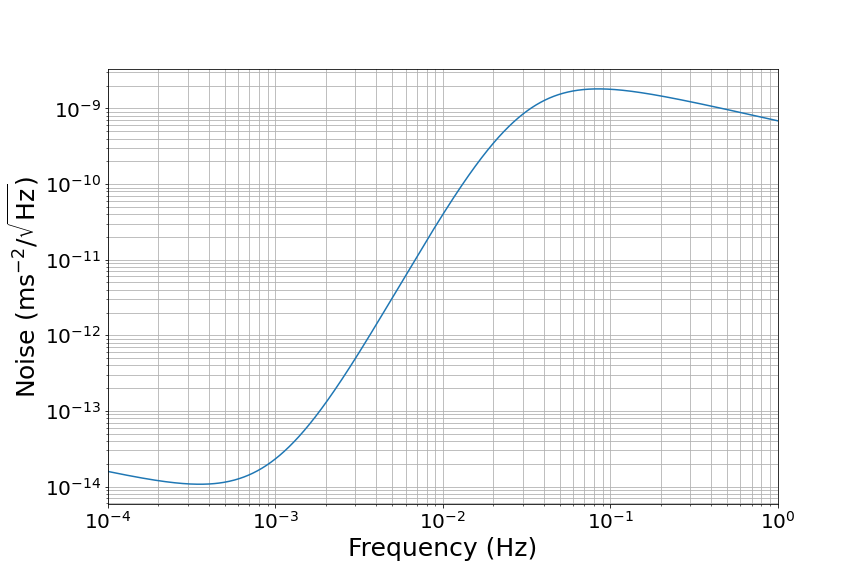}
            \caption{Modelled acceleration noise based on LISA Path Finder data}
            \label{fig:Acceleration Noise}
    \end{figure}

Finally, we model the spacecraft motion (Fig~\ref{fig:Acceleration Noise}) to conservatively fit  the residual acceleration noise ASD of the LISA Pathfinder mission (grey trace in Fig~10 of \cite{PhysRevD.99.082001}):
    \begin{equation}
            S_a^{1/2}(f)\lesssim 6.87\times 10^{-10} \left(\frac{\sqrt{1+\left(\frac{1\,{\rm mHz}}{f}\right)^2}}{\sqrt{1+\left(\frac{30\,{\rm mHz}}{f}\right)^2}}\right)^{4.5}\left(\frac{1\,{\rm Hz}}{f}\right)^{0.5}\frac{ {\rm ms}^{-2}}{\sqrt{{\rm Hz}}}\\
    \end{equation}    

The corresponding frequency noise is given by:
    \begin{equation}
        \phi_{{\rm X}_{ij}}(f)= \frac{ S_a^{1/2}(f)}{2\pi f\lambda}\frac{{\rm Hz}}{\sqrt{{\rm Hz}}}
    \end{equation}
    where $\lambda$ is the laser wavelength.




\subsection{Errors in estimates of initial Doppler shift and its derivatives }

We follow the Doppler measurement concept used by \cite{Kirk} to estimate the initial Doppler shift and its derivatives.  The error in these measurements for a given LISA arm, is the standard deviation of the measured beat frequency between the local  and the transponded light from the far spacecraft and its derivatives. We use 3 stages of averaging for estimating the errors zeroth, first, and second Doppler derivative measurements. These expressions for an averaging window size of $T~{\rm seconds}$ are given by :
\begin{eqnarray}
    \nu_{0_{ij}}=&2\sqrt{\int_{0}^{\infty}\phi_{{\rm L}1}^2(f){\rm sinc}^6(fT)\sin ^2 (\pi f \tau_{ij})df}\\
    \gamma_{0_{ij}}=&2\sqrt{\int_{0}^{\infty}(2\pi f)^2\phi_{{\rm L}1}^2(f){\rm sinc}^6(fT)\sin ^2 (\pi f \tau_{ij})df}\\
    \alpha_{0_{ij}}=&2\sqrt{\int_{0}^{\infty}(2\pi f)^4\phi_{{\rm L}1}^2(f){\rm sinc}^6(fT)\sin ^2 (\pi f \tau_{ij})df}
\end{eqnarray}
The corresponding common and differential Doppler errors for averaging windows of $T=200~ {\rm seconds  }$ and $T=40000 ~{\rm seconds }$ are given in table \ref{tab:Initial errors in Doppler estimates and its derivatives}.\\
One caveat when estimating the Doppler error is that if the estimated error made by the measurement is greater than the Doppler estimate given by the orbit simulation \cite{Orbit}, we do not subtract the measured Doppler estimate. The residual Doppler error is then taken to be the value given by the orbit data (denoted by * in table \ref{tab:Initial errors in Doppler estimates and its derivatives}). 


\begin{table}[H]
\begin{center}
\begin{tabular}{|c|c|c|}
\hline
&\multicolumn{2}{c|}{Averaging Time Window T}\\
\hline
&200 {\rm  seconds}& 40000 {\rm  seconds} \\ 
\hline
 $\nu_{0+}$ & $6.88~{\rm Hz}$ &$4.89 ~{\rm mHz}$ \\
 $\nu_{0-}$& $-0.04~{\rm Hz}$& $-2.89\times 10^{-2}\,{\rm mHz}$\\
 $\gamma_{0+}$ &  13.08 {\rm mHz/s} & 1.96~$\mu${\rm Hz/s}\\
 $\gamma_{0-}$& $7.71\times 10^{-2}~{\rm mHz/s}$& $-1.15\times 10^{-2}\mu{\rm Hz/s}$\\
$\alpha_{0+}$ &$-0.24~\mu{\rm Hz}/s^2*$ &$8.43\times 10^{-1}~{\rm nHz}/s^2$ \\
$\alpha_{0-}$ & $-0.12~\mu{\rm Hz}/s^2*$&$-4.98 {\rm pHz}/s^2$ \\
\hline
\end{tabular}
\caption{Initial errors in Doppler estimates and its derivatives for a  laser prestabilized to the LISA requirement, using  $200$ seconds and $40000$ seconds averaging windows}
\label{tab:Initial errors in Doppler estimates and its derivatives}
\end{center}
\end{table}


\subsection{Controller poles and zeros}
\label{subsec:Controllerpolesandzeros}
The arm locking controller takes the parametric form:
\begin{equation}
\fl    G(s)=\frac{1}{2}\left(\frac{g_1 s}{s+p_1}\right)^3 \left(\frac{g_2 s}{s+p_2}\right)\left(\frac{g_3(s+z_3)}{s+p_3}\right)^5\left(\frac{g_4}{(s+p_{41})(s+p_{42})}+\sum_{k=1}^{9}\frac{g_{5k}}{s+p_{5k}}\right)
\label{eqn:Armlockingcontroller}
\end{equation}

The arm locking open loop response is given in terms of the arm sensor, controller and system delays ($\tau_{\rm sys}=17.6~\mu{\rm s} $ \cite{Kirk})by:
\begin{equation}
    G_{\rm O}(s)=G(s)P_{\rm M}(s)\exp(-is\tau_{\rm sys})
\end{equation}
The response  essentially has four high pass filters for AC coupling  below the LISA band followed by a transition to a high constant gain phase $\mathcal{O}(10^5) $ in the LISA science band and a $f^{-0.66}$ roll off at frequencies beyond $\mathcal{O} (1~{\rm Hz})$. Additionally,
the plot shows that we are able to maintain a phase margin of $30^\circ$ from $180^\circ$ at unity gain frequencies below $\approx 5 \,{\rm kHz}$ which defines our arm locking bandwidth.

\begin{figure}[H]
\centering
\includegraphics[width=\textwidth]{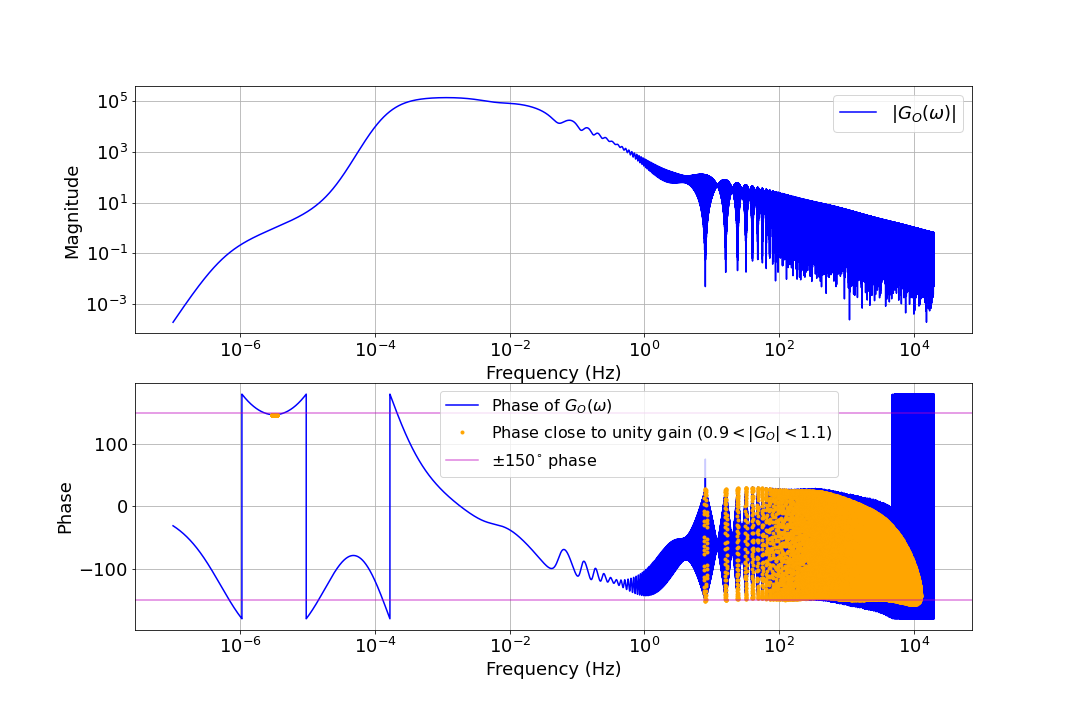}
\caption{Modified dual arm open loop Bode plot}     
 \label{fig:MDAL open loop Bode plot}    
\end{figure}
Comparing our controller poles and zeros to the ones used in \cite{Kirk}, we see that the higher frequency poles, i.e, $(p_3,p_{41},p_{42},p_{51}\ldots p_{59})$  have been scaled by a factor of 2 to account for the arm lengths being half of the original LISA mission. The amplitude and decay time  Doppler frequency  pulling transient  (Fig. \ref{fig:Doppler frequency pulling transients} )is characterized by  the four high pass filters, the lower unity gain frequency and the zero of the band pass filters in Eq. \ref{eqn:Armlockingcontroller} (i.e., by $p_1,p_2,p_3,z_3,{\rm  and }f_{{\rm UG}}$).\\\\

Let $\tau_{decay}$ represent the characteristic decay time of the Doppler pulling transient and $A_{n}$ represent the magnitude of the transient contribution by the $n^{{\rm th}}$ Doppler derivative.
It can be shown that for the same initial Doppler error estimates, the transformation \begin{equation}
    (p_1,p_2,p_3,z_3,f_{{\rm UG}})\longrightarrow \beta(p_1,p_2,p_3,z_3,f_{{\rm UG}})
\end{equation}
leads to 
\begin{eqnarray}
    \tau_{\rm decay}&\longrightarrow \tau_{\rm decay}/\beta\\
    A_{n}& \longrightarrow A_{n}/\beta^{n+1}
\end{eqnarray}\\
Thus, while  scaling down these poles and zeros increases the controller gain, it  worsens the Doppler frequency pulling characteristics. With this in mind we chose to scale  $(p_1,p_2,p_3,z_3,f_{{\rm UG}})$ by  $\beta=2/3$. This choice ensures that the arm locking output (eqn\ref{eqn:armlockloopoutallnoises}) is not limited by the controller gain in the $3\,{\rm mHz}-300\,{\rm mHz}$ band.
\begin{table}[H]
\begin{center}
\begin{tabular}{|c|c|c|}
\hline
     Zeros (Hz) & Poles (Hz) & Gain  \\
     \hline
     $z_1=0$ & $p_1=2\pi\times1.6\times 10^{-7}$ & $g_1=1$\\
     \hline
     $z_2=0$ & $p_2=2\pi\times  420\times 10^{-6}$ & $g_{2}=\left(\frac{0.95}{2\pi f_{{\rm UG}}}p_2\right)$\\
     &&$f_{\rm UG}=3.2\,\mu{\rm Hz}$\\
     \hline
     $z_3=2\pi\times 73.2\times 10^{-6}$ & $p_{3}=2\pi\times 370\times 10^{-6}$ & $g_3=p_3/z_3$\\
     \hline
     & $p_{41}=2\pi\times 6\times 10^{-3}$
     & $g_4=p_{41}p_{42}$\\
     &$p_{42}=2\pi\times 576 \times 10^{-3}$&\\
     \hline
           & $p_{51}=2\pi\times 6\times 10^{-3}$ & $g_{51}=1.3\times 10^{-3}$\\
           & $p_{52}=2\pi\times 6\times 10^{-2}$ & $g_{52}=3.7\times 10^{-3}$\\
           & $p_{53}=2\pi\times 6\times 10^{-1}$ & $g_{53}=4.2\times 10^{-3}$\\
           & $p_{54}=2\pi\times 6$  & $g_{54}=16\times 10^{-3}$\\
           & $p_{55}=2\pi\times 6\times 10^{1}$ & $g_{55}=30\times 10^{-3}$\\
           & $p_{56}=2\pi\times 6\times 10^{2}$ & $g_{56}=69\times 10^{-3}$\\
           & $p_{57}=2\pi\times 6\times 10^{3}$ & $g_{57}=0.11$\\
           & $p_{58}=2\pi\times 6\times  10^{4}$ & $g_{58}=0.33$\\
           & $p_{59}=2\pi\times 6\times 10^{5}$ & $g_{59}=0.7$\\
     \hline
  
\end{tabular}
\caption{Modified dual arm locking controller design}
\label{tab:MDAL_controller}
\end{center}
\end{table}




\section{Modified dual arm locking noise performance for the new LISA mission}
\label{sec:Modified dual arm locking noise performance for the new LISA mission}

\begin{figure}[H]
    \centering
    \begin{minipage}{0.49\textwidth}
        \includegraphics[width=\textwidth]{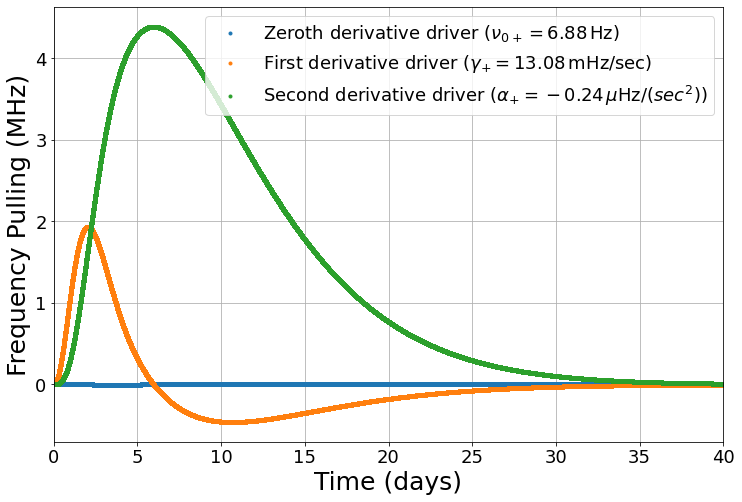}
    \end{minipage}
    \hfill
    \begin{minipage}{0.5\textwidth}
        \includegraphics[width=\textwidth]{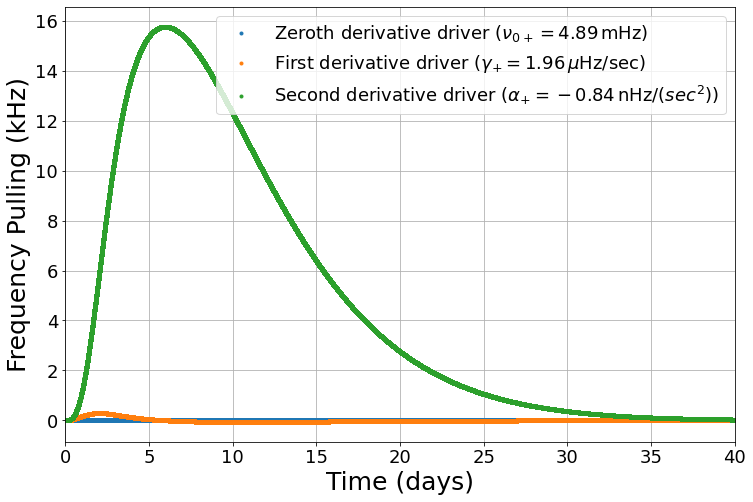}
    \end{minipage}
    \caption{Doppler frequency pulling of laser pre-stabilized  to the LISA requirement (Eq.\ref{eqn:Lisareq}) with $T=200$  seconds (left) and $T=40000$ seconds (right)  averaging windows. The three traces correspond to contributions from the  errors in the zeroth (blue), first (orange) and second (green) Doppler derivative estimation.}
    \label{fig:Doppler frequency pulling transients}
\end{figure}

We now have all the tools to calculate  the Doppler pulling and the noise suppression achieved by the modified dual arm locking control loop. Figure \ref{fig:Doppler frequency pulling transients} shows the Doppler frequency pulling transients for initial Doppler measurements made with $\,200 \,{\rm s}\, {\rm  and } \,40000 \,{\rm  s}$ averaging windows. We see that for $200\, {\rm  s}$ averaging window, the control system is required to accommodate a laser frequency to drift of about $4.5~{\rm MHz}$ in 5 days. Increasing the Doppler estimation averaging to $40000\,{\rm s}$  relaxes this requirement to about $16 \,{\rm kHz}$ in 5 days. Significant improvement in the Doppler pulling  performance is seen  once the averaging window is large enough $\mathcal{O}(10^4)$ seconds to make the Doppler measurement of the second derivative ($\alpha_{0+}$) more precise than the value given by the orbit model.
 
\begin{figure}[H]
    \centering
    \includegraphics[scale=0.25]{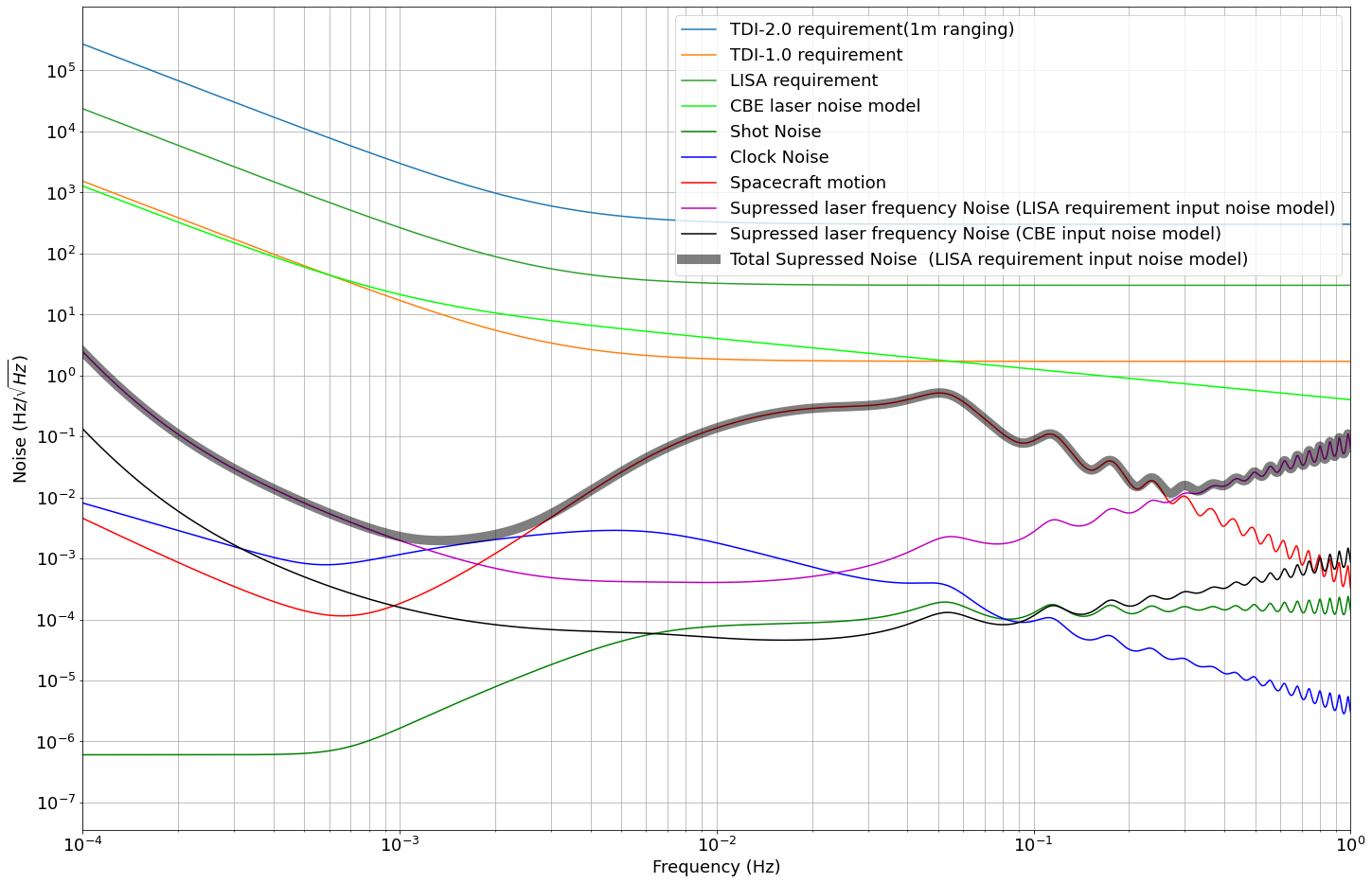}
\caption{Modified dual arm locking noise performance with $\Delta \tau =62 \,{\rm  ms}$ (Eq.\ref{eqn:armlockloopoutallnoises}). The total suppressed noise is plotted along with the individual contributions from the residual laser frequency noise, clock noise, shot noise and spacecraft motion. All traces are compared with TDI requirements, LISA requirements and the CBE cavity stabilized laser.}
\label{fig:MDALvsTDIreqsandinputlasernoise}
\end{figure}

The corresponding modified dual arm locking noise performance is plotted in Fig. \ref{fig:MDALvsTDIreqsandinputlasernoise}. We see that the arm locking stabilized laser frequency noise is far better than the CBE  and also meets the requirement for TDI 1.0. With the input laser noise prestabilized to the LISA requirement (Eq.\ref{eqn:Lisareq}), we are limited by spacecraft motion in the $3\,{\rm mHz}<f<300\,{\rm mHz}$ frequency band , clock noise in the $1\,{\rm mHz}<f<2\,{\rm mHz}$ band and by residual laser frequency noise in the $f<1\,{\rm mHz}$ and $f>300\,{\rm mHz}$  bands. The noise suppression ranges from $\mathcal{O}(10^4)$ in the $0.1 ~{\rm mHz}<f<4 ~{\rm mHz}$  frequency band to $\mathcal{O}(10^2)$ in the $f>10~{\rm mHz}$ frequency band. Additionally, with the CBE cavity stabilized laser input, we are only gain limited at frequencies $f<0.3\,{\rm mHz}$ .

            
     

\begin{figure}[H]
    \centering
    \begin{minipage}{0.48\textwidth}
        \includegraphics[width=\textwidth]{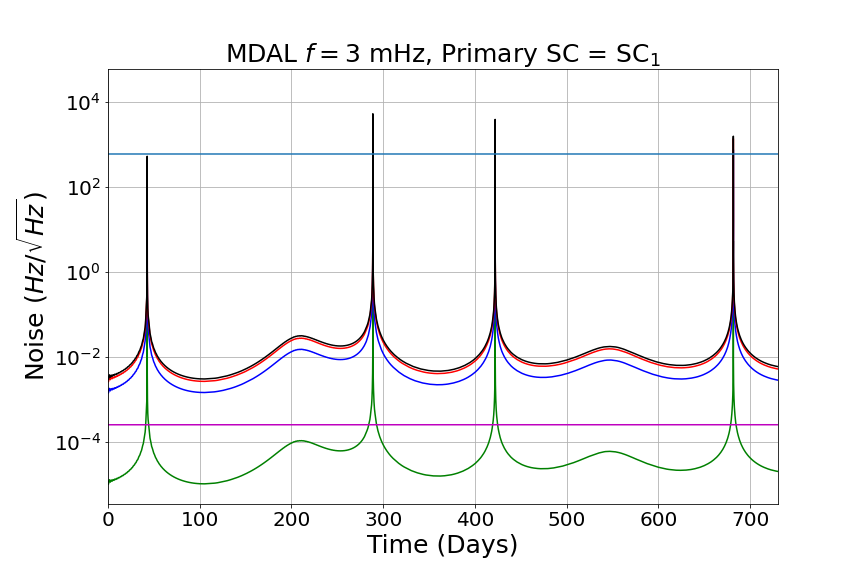}
    \end{minipage}
    \hfill
    \begin{minipage}{0.48\textwidth}
        \includegraphics[width=\textwidth]{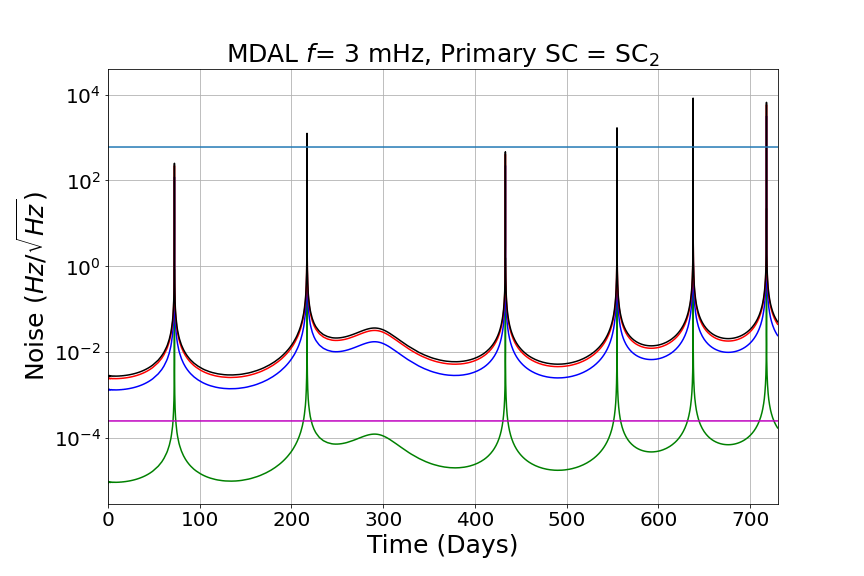}
    \end{minipage}
    \begin{minipage}{0.48\textwidth}
        \includegraphics[width=\textwidth]{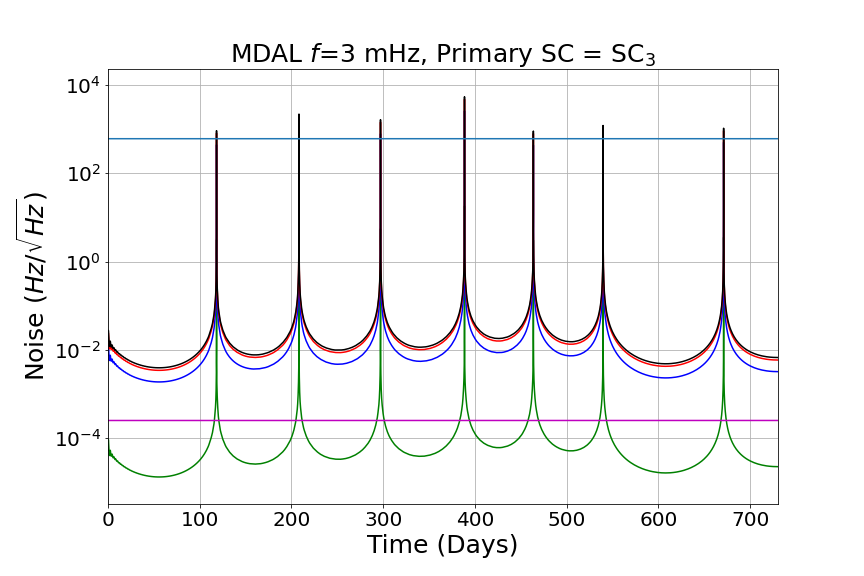}
    \end{minipage}
    \hfill
    \begin{minipage}{0.48\textwidth}
        \includegraphics[width=\textwidth]{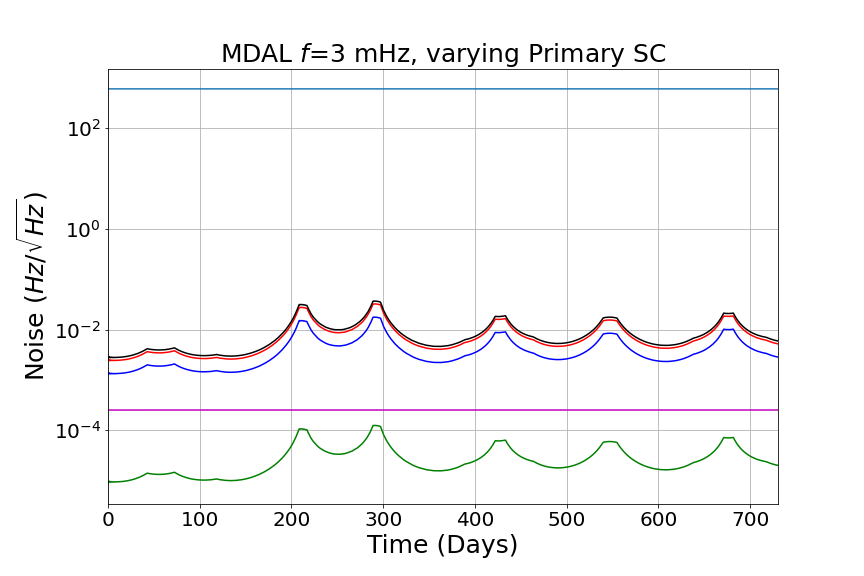}
    \end{minipage}  
    \caption{Residual laser frequency noise (magenta), spacecraft motion (red), clock noise 
    (blue), shot noise (green) and the cumulative noise (black) performance of  modified dual arm locking sensors measured  at 3 mHz over the first 2 years of the new LISA mission. The plots in the top panel and the bottom left one correspond to  
    SC$_1$, SC$_2$, SC$_3$ being  the primary spacecraft respectively. The  bottom right plot is the performance obtained by enabling the switch of primary spacecraft for optimum $\Delta \tau$ values. All curves are compared to the TDI 2.0 requirement at 3 mHz (light blue).}
    \label{fig:MDAL2yrs}
\end{figure}

Using the the LISA orbit data, we also plot the modified dual arm locking noise curves for 2 years of the new LISA mission at $f= 3~{\rm mHz}$ in Fig.~\ref{fig:MDAL2yrs}. The top panel and the bottom left plot are plots corresponding to the 3 cases of  SC1, SC2 and SC3 being the primary spacecraft respectively. At this frequency, we are dominated by the spacecraft motion at $10^{-2} {\rm Hz}/\sqrt{{\rm Hz}}$ for a significant part of the mission. This, similar to Fig.~\ref{fig:MDALvsTDIreqsandinputlasernoise}, corresponds to a noise suppression of $\mathcal{O}(10^4)$ from the $30~{\rm Hz}/\sqrt{{\rm Hz}}$ pre-stabilized input laser frequency noise model. The only times at which the noise level goes above the TDI requirement are times at which $\Delta \tau \approx 0$. Physically, these are times at which the differential arm has zero response, which forces a singularity in the integrator $(\frac{1}{s\Delta \tau})$ used to generate the arm response (Eq. \ref{eqn:diffintegrator})and hence causes large amplifications in clock noise, shot noise and spacecraft motion terms.
We see that these times are different for the 3 different configurations. Hence, as shown in the bottom right plot of Fig.~\ref{fig:MDAL2yrs}, allowing the switching of the primary spacecraft will enable us to maximize the value of $\Delta \tau $ and hence optimize the noise performance.



\section{Conclusion}\label{sec:conclusion}

To summarize, following  the footsteps of McKenzie et.al \cite{Kirk}, we have estimated the modified dual arm locking noise performance for the new LISA mission. The new developments compared to \cite{Kirk} are the shorter arms, improved new clock noise levels and a new model for the spaceraft motion based on LISA Pathfinder data. In this work, we assume a cavity stabilized input laser noise and we tweak the controller design proposed by \cite{Kirk} to have more gain in the LISA band, by lowering the lower unity gain frequency. Doing so does worsen the Doppler frequency pulling rate but this new frequency pulling rate is still within the acceptable bounds of laser tuning range. In order to significantly improve the Doppler pulling characteristics at lock acquisition without sacrificing the gain in the LISA band, we also propose to increase the averaging time window for the initial Doppler estimate from $T=200\,{\rm seconds}$ to $T\approx 40000 \,{\rm  seconds}$.\\\\

\section{Acknowledgement}

This work is supported by NASA grant 80NSSC20K0126.

\section*{References}
\bibliographystyle{unsrt}
\bibliography{Armlocking.bib}

\begin{thebibliography}{10}

\bibitem{L3proposal}
Pau Amaro-Seoane et~al.
\newblock Laser interferometer space antenna, 2017.

\bibitem{LIGOScientific:2014pky}
J.~Aasi et~al.
\newblock {Advanced LIGO}.
\newblock {\em Class. Quant. Grav.}, 32:074001, 2015.

\bibitem{LIGOScientific:2007fwp}
B.~P. Abbott et~al.
\newblock {LIGO: The Laser interferometer gravitational-wave observatory}.
\newblock {\em Rept. Prog. Phys.}, 72:076901, 2009.

\bibitem{VIRGO:2012dcp}
T.~Accadia et~al.
\newblock {Virgo: a laser interferometer to detect gravitational waves}.
\newblock {\em JINST}, 7:P03012, 2012.

\bibitem{VIRGO:2014yos}
F.~Acernese et~al.
\newblock {Advanced Virgo: a second-generation interferometric gravitational
  wave detector}.
\newblock {\em Class. Quant. Grav.}, 32(2):024001, 2015.

\bibitem{PhysRevD.88.043007}
Yoichi Aso, Yuta Michimura, Kentaro Somiya, Masaki Ando, Osamu Miyakawa,
  Takanori Sekiguchi, Daisuke Tatsumi, and Hiroaki Yamamoto.
\newblock Interferometer design of the {KAGRA } gravitational wave detector.
\newblock {\em Phys. Rev. D}, 88:043007, Aug 2013.

\bibitem{akutsu2020overview}
T.~Akutsu et~al.
\newblock Overview of {KAGRA}: Detector design and construction history, 2020.

\bibitem{Grote:2010zz}
H.~Grote.
\newblock {The GEO 600 status}.
\newblock {\em Class. Quant. Grav.}, 27:084003, 2010.

\bibitem{Dooley_2016}
K~L Dooley et~al.
\newblock {GEO} 600 and the {GEO}-{HF} upgrade program: successes and
  challenges.
\newblock {\em Classical and Quantum Gravity}, 33(7):075009, mar 2016.

\bibitem{Kroker:2015pmg}
Stefanie Kroker and Ronny Nawrodt.
\newblock {The Einstein telescope}.
\newblock {\em IEEE Instrum. Measur. Mag.}, 18(3):4--8, 2015.

\bibitem{reitze2019cosmic}
David Reitze, Rana~X Adhikari, Stefan Ballmer, Barry Barish, Lisa Barsotti,
  GariLynn Billingsley, Duncan~A. Brown, Yanbei Chen, Dennis Coyne, Robert
  Eisenstein, Matthew Evans, Peter Fritschel, Evan~D. Hall, Albert Lazzarini,
  Geoffrey Lovelace, Jocelyn Read, B.~S. Sathyaprakash, David Shoemaker, Joshua
  Smith, Calum Torrie, Salvatore Vitale, Rainer Weiss, Christopher Wipf, and
  Michael Zucker.
\newblock Cosmic {E}xplorer: The {U.S.} contribution to gravitational-wave
  astronomy beyond {LIGO}, 2019.

\bibitem{10.1007/s41114-020-00029-6}
Massimo Tinto and Sanjeev~V Dhurandhar.
\newblock {Time-delay interferometry}.
\newblock {\em Living Reviews in Relativity}, 24(1):1, 2021.

\bibitem{PhysRevD.65.082003}
Massimo Tinto, F.~B. Estabrook, and J.~W. Armstrong.
\newblock Time-delay interferometry for {LISA}.
\newblock {\em Phys. Rev. D}, 65:082003, Apr 2002.

\bibitem{freqplanning}
{LISA} performance model and error budget, {LISA-LCST-INST-TN-003}.
\newblock 2018.

\bibitem{2017}
B~Bachman, G~de~Vine, J~Dickson, S~Dubovitsky, J~Liu, W~Klipstein, K~McKenzie,
  R~Spero, A~Sutton, B~Ware, and C~Woodruff.
\newblock Flight phasemeter on the laser ranging interferometer on the {GRACE}
  follow-on mission.
\newblock 840:012011, may 2017.

\bibitem{Sheard:2003fz}
B.~S. Sheard, M.~B. Gray, D.~E. McClelland, and D.~A. Shaddock.
\newblock {Laser frequency stabilization by locking to a LISA arm}.
\newblock {\em Phys. Lett. A}, 320:9--21, 2003.

\bibitem{Kirk}
Kirk McKenzie, Robert~E. Spero, and Daniel~A. Shaddock.
\newblock Performance of arm locking in lisa.
\newblock {\em Phys. Rev. D}, 80:102003, Nov 2009.

\bibitem{Yinan}
Yiu Yinan.
\newblock {\em Arm Locking for Laser Interferometer Space Antenna}.
\newblock PhD thesis, University of Florida, 2011.

\bibitem{Sutton}
Andrew Sutton and Daniel~A. Shaddock.
\newblock Laser frequency stabilization by dual arm locking for {LISA}.
\newblock {\em Phys. Rev. D}, 78:082001, Oct 2008.

\bibitem{Orbit}
{LISA} frequency planning, {LISA- AEI-INST-TN-002}.
\newblock 2018(unpublished).

\bibitem{Clocknoisereference}
Massimo Tinto and Olaf Hartwig.
\newblock Time-delay interferometry and clock-noise calibration.
\newblock {\em Phys. Rev. D}, 98:042003, Aug 2018.

\bibitem{PhysRevD.99.082001}
M.~Armano, H.~Audley, J.~Baird, P.~Binetruy, M.~Born, D.~Bortoluzzi,
  E.~Castelli, A.~Cavalleri, A.~Cesarini, A.~M. Cruise, K.~Danzmann,
  M.~de~Deus~Silva, I.~Diepholz, G.~Dixon, R.~Dolesi, L.~Ferraioli, V.~Ferroni,
  E.~D. Fitzsimons, M.~Freschi, L.~Gesa, F.~Gibert, D.~Giardini, R.~Giusteri,
  C.~Grimani, J.~Grzymisch, I.~Harrison, G.~Heinzel, M.~Hewitson,
  D.~Hollington, D.~Hoyland, M.~Hueller, H.~Inchausp\'e, O.~Jennrich,
  P.~Jetzer, N.~Karnesis, B.~Kaune, N.~Korsakova, C.~J. Killow, J.~A. Lobo,
  I.~Lloro, L.~Liu, J.~P. L\'opez-Zaragoza, R.~Maarschalkerweerd, D.~Mance,
  N.~Meshksar, V.~Mart\'{\i}n, L.~Martin-Polo, J.~Martino, F.~Martin-Porqueras,
  I.~Mateos, P.~W. McNamara, J.~Mendes, L.~Mendes, M.~Nofrarias, S.~Paczkowski,
  M.~Perreur-Lloyd, A.~Petiteau, P.~Pivato, E.~Plagnol, J.~Ramos-Castro,
  J.~Reiche, D.~I. Robertson, F.~Rivas, G.~Russano, J.~Slutsky, C.~F. Sopuerta,
  T.~Sumner, D.~Texier, J.~I. Thorpe, D.~Vetrugno, S.~Vitale, G.~Wanner,
  H.~Ward, P.~J. Wass, W.~J. Weber, L.~Wissel, A.~Wittchen, and P.~Zweifel.
\newblock {LISA} pathfinder platform stability and drag-free performance.
\newblock {\em Phys. Rev. D}, 99:082001, Apr 2019.

\end{thebibliography}

\end{document}